\documentclass[secnumarabic,amssymb, nobibnotes, aps, pre, twocolumn]{revtex4-1}

\usepackage[english]{babel}
\usepackage{amsmath}
\usepackage{graphicx}  
\usepackage[caption=false]{subfig}
\usepackage{color}
\usepackage{tikz}
\usetikzlibrary{chains}
\usetikzlibrary{arrows}
\usetikzlibrary{external}
\tikzset{external/system call={latex \tikzexternalcheckshellescape -halt-on-error
-interaction=batchmode -jobname "\image" "\texsource";
dvips -o "\image".ps "\image".dvi;
ps2eps "\image.ps"}}
\begin{document}

\title{Extended Parrondo's Game and Brownian Ratchets: Strong and Weak Parrondo Effect}

\author{Degang Wu}\author{Kwok Yip Szeto}
\email[Corresponding author ]{phszeto@ust.hk}
\affiliation{Department of Phyiscs,\\
The Hong Kong University of Science and Technology,\\
Clear Water Bay, Hong Kong, HKSAR, China
}
\date{\today}
\begin{abstract}
Inspired by the flashing ratchet, Parrondo's game presents an apparently paradoxical situation. Parrondo's game consists of two individual games, game A and game B. Game A is a slightly losing coin-tossing game. Game B has two coins, with an integer parameter $M$. If the current cumulative capital (in discrete unit) is a multiple of $M$, an unfavorable coin $p_b$ is used, otherwise a favorable $p_g$ coin is used. Paradoxically, combination of game A and game B could lead to a winning game, which is the Parrondo effect. We extend the original Parrondo's game to include the possibility of $M$ being either $M_1$ or $M_2$. Also, we distinguish between strong Parrondo effect, i.e. two losing games combine to form a winning game, and weak Parrondo effect, i.e. two games combine to form a better-performing game. We find that when $M_2$ is not a multiple of $M_1$, the combination of $B(M_1)$ and $B(M_2)$ has strong and weak Parrondo effect for some subsets in the parameter space $(p_b,p_g)$, while there is neither strong nor weak effect when $M_2$ is a multiple of $M_1$. Furthermore, when $M_2$ is not a multiple of $M_1$, stochastic mixture of game A may cancel the strong and weak Parrondo effect. Following a discretization scheme in the literature of Parrondo's game, we establish a link between our extended Parrondo's game with the analysis of discrete Brownian ratchet. We find a relation between the Parrondo effect of our extended model to the macroscopic bias in a discrete ratchet. The slope of a ratchet potential can be mapped to the fair game condition in the extended model, so that under some conditions, the macroscopic bias in a discrete ratchet can provide a good predictor for the game performance of the extended model. On the other hand, our extended model suggests a design of a ratchet in which the potential is a mixture of two periodic potentials.
\end{abstract}

\maketitle

\section{Introduction}

\makeatletter{}In 1992, Ajdari and Prost discovered a Brownian ratchet mechanism \cite{ajdari_mouvement_1992}, which was named by Astumian and Bier \cite{astumian_fluctuation_1994} the flashing ratchet. Inspired by the flashing ratchet, Parrondo \cite{parrondo_how_1996} invented the games of chance later known as Parrondo's games, in which two losing games can be combined following a random or periodic strategy leading to a winning game. Later, Allison et al. \cite{allison_physical_2002-1} and Toral et al. \cite{toral_parrondos_2003} demonstrated that Parrondo's game can be described by a discrete Fokker-Planck equation, thus a more rigorous relation between Parrondo's game and Brownian ratchet was established. From the perspective of game, the optimal sequence for a given set of parameters for Parrondo's games was discovered by Dinis \cite{dinis_optimal_2008}. 

The games have also received attention in many other fields \cite{abbott_asymmetry_2010}, ranging from Brownian ratchets \cite{lee_minimal_2003,astumian_paradoxical_2005}, nonlinear dynamics \cite{kocarev_lyapunov_2002-1,arena_game_2003,almeida_can_2005,danca_parrondos_2012}, biology \cite{wolf_diversity_2005,reed_two-locus_2007}, {chemistry \cite{osipovitch_systems_2009}}, and economics \cite{spurgin_switching_2005}. Different variants of the original Parrondo's games have been developed, including history-dependent Parrondo's game \cite{parrondo_new_2000}, Parrondo's game with self-transition \cite{allison_physical_2002-1} and multi-player version of Parrondo's game \cite{dinis_optimal_2003,dinis_inefficiency_2004,parrondo_brownian_2004}. In particular, a variant called Parrondo's game with one-dimensional spatial dependence \cite{toral_cooperative_2001} has been investigated by Mihailovic \cite{mihailovic_one_2003} and generalized to synchronous case \cite{mihailovic_synchronous_2003} and two-dimensional case \cite{mihailovic_cooperative_2006}. Whether scale free network allows Parrondo's games with spatial dependence was also investigated \cite{toyota_does_2012,toyota_parrondo_2012,toyota_second_2012}. Quantum versions of Parrondo's games have also received attention \cite{chen_quantum_2010, bulger_position-dependent_2008, bleiler_properly_2011, pawela_cooperative_2013}. An optical model of quantum Parrondo's game was implemented experimentally \cite{si_optical_2012}, based on the techniques developed in Ref.~\cite{schmid_experimental_2010}. In an interesting paper by Harmer et al.~\cite{harmer_brownian_2001}, the authors discussed several open questions about Parrondo games. One of these open questions concerned the possibility of different $M$ during play.

Since Parrondo's game was inspired by the flashing ratchet, a question was raised whether one can infer characteristics of certain continuous Brownian ratchets from extended versions of the original Parrondo's game. This line of research was pursued by Harmer et al. \cite{harmer_parrondos_2000} with preliminary results. In addition to the usual game A and game B, the integer parameter $M$ in the B game can assume different values between 3 and 10 with equal probability at each game. It was demonstrated by simulations that under this setting other counter-intuitive phenomena would occur. The motivation for randomizing $M$ was that $M$ controls the period of the ratchet potential and therefore randomizing $M$ means randomizing the period of the ratchet potential. The Parrondo's game extended in this way corresponds to a type of Brownian ratchets other than the flashing ratchet. We follow this line of research and use a different but similar model, in which $M$ can be either of $M_1$ and $M_2$. Restricting $M$ to be one of only two values allows systematic investigations while one can still observe interesting phenomena. Among the various properties of our extended model, we point out the significance of weak Parrondo effect, which is the situation when two games, which need not be both losing, combines to form a better game in the sense of losing less or even winning more. Weak Parrondo effect is a natural and meaningful extension to the well-known Parrondo effect: two losing games combine to become a winning game. We show that this distinction between the strong and weak Parrondo effect is significant in our extended model.

The paper is organized as follows: we begin by introducing the original Parrondo's game and its Markov chain formulation in Sec. II. In Sec. III we present the formulation of our extended Parrondo's game. In Sec. IV, we show the conditions under which there are strong Parrondo effects for our extended Parrondo game with $B(M_1)$ and $B(M_2)$ and the further mixture with game A. In Sec. V, we discuss new features of mixing two B games (with or without mixing also with game A) that we call the weak Parrondo effect. In Sec. VI, we apply the Fokker-Planck dicretization scheme on the extended model, and show the properties of our extended model from the perspective of discrete Brownian ratchets. Concluding remarks can be found in Sec. VII. 

\section{The Original Parrondo's Game: (A,B($M$))}

\makeatletter{}The original Parrondo's game consists of two individual coin tossing games, namely game A and game B. Game A has only one coin, whose winning probability is $p_A=1/2-\epsilon$, where $\epsilon$ is a small and positive number. Let $X(t)$ be the cumulative capital of the player at time $t$, a non-negative integer. If the player keeps playing game A, the average capital satisfies
\begin{equation}
\langle X(t+1)\rangle=\langle X(t)\rangle+2p_A-1,
\end{equation}
where $\langle\cdot\rangle$ is understood as ensemble average. We define the long-term expected gain as
\begin{equation}
g\equiv\lim_{t\rightarrow\infty}\langle X(t+1)\rangle - \langle X(t)\rangle,
\label{eq:expected_gain}
\end{equation}
which in many cases exists. If $\langle X(t+1)\rangle - \langle X(t)\rangle$ oscillates in a limit cycle, then $g$ is understood to be an average over a limit cycle.
Thus with $p_A=1/2$, $g$ is zero as this is a trivial unbiased random walk. In the context of Parrondo's game \cite{nowak_two_2005}, a winning game is one that has positive $g$. A fair game is one with $g=0$ in the long run or $g$ with zero average over a limit cycle. For positive $\epsilon$, $g=-2\epsilon$ and game A is a losing game

Game B has two coins, one ``good" coin and one ``bad" coin. B game has an integer parameter $M$. If $X(t)$ is a multiple of $M$, then $X(t+1)$ is determined by the ``bad" coin with winning probability $p_b=1/10-\epsilon$, otherwise the ``good" coin with winning probability $p_g=3/4-\epsilon$ is used.

Similar to game A, if the player keeps playing game B only, the average capital satisfies
\begin{equation}
\langle X(t+1)\rangle = \langle X(t)\rangle+2(\pi_0(t)p_b+(1-\pi_0(t))p_g)-1,
\end{equation}

which explicitly depends on $\pi_0$, the probability that $X(t)=0\mod M$. Harmer and Abbott \cite{harmer_game_1999} showed that  game B is a losing game with $p_b=1/10-\epsilon$, $p_g=3/4-\epsilon$ and $M=3$, with positive $\epsilon$.

If we model the Parrondo's game as a discrete-time Markov chain as in Ref.~\cite{harmer_review_2002}, we can define the probability vector (for simplicity we set $M=3$ for the purpose of demonstration) $\boldsymbol{\pi}(t)\equiv\left(\pi_0(t),\pi_1(t),\pi_2(t)\right)^T$. Accordingly, the transition matrix for game A is
\begin{equation}
\Pi_A=
\begin{pmatrix}
0 & 1-p_A & p_A \\
p_A & 0 & 1-p_A \\
1-p_A & p_A & 0 \\
\end{pmatrix},
\end{equation}
such that the time evolution equation is $\boldsymbol{\pi}(t+1)=\Pi_A\boldsymbol{\pi}(t)$. Similarly, the transition matrix for game B is
\begin{equation}
\Pi_B=
\begin{pmatrix}
0 & 1-p_g & p_g \\
p_b & 0 & 1-p_g \\
1-p_b & p_g & 0 \\
\end{pmatrix}.
\end{equation}

The stochastic mixture of game A and B has the following transition matrix
\begin{equation}
\Pi=\gamma \Pi_A+(1-\gamma)\Pi_B,
\end{equation}
where $\gamma$ is the probability of playing game A in the stochastic mixing of game A and B.
Parrondo's game can also be played according to a periodic game sequence such as ABABB, in which case the probability vector is evolved by multiplying $\boldsymbol{\pi}$ with $\Pi_A$ or $\Pi_B$ according to the sequence.

Parrondo's game has a seemingly paradoxical property that while game A and B are losing when they are played individually, the stochastic mixture of game A and B, or playing according to a deterministic sequence may lead to a winning combined game for small positive value of $\epsilon$. For the detailed analysis of the apparent paradox, please refer to Ref.~\cite{harmer_review_2002}. In summary, since the two games are coupled non-linearly through $X(t)$, the combination of the two losing games are non-linear and in general it is not surprising that a winning game can emerge from their combination. In the context of Parrondo's game, the phenomenon that two losing games can be combined to produce a winning game is called the Parrondo effect. An interesting and related phenomenon also deserves investigation, namely that two games, not necessarily losing, combine to form a game that performs better, though not necessarily winning, than either of the two individual games, which will be called weak Parrondo effect. Obviously, the criteria for Parrondo effect fits the criteria for weak Parrondo effect, but the reverse is not true.  

\section{The Extended Game: ($B(M_1)$, $B(M_2)$)}

\begin{figure}
\includegraphics[width=0.8\columnwidth]{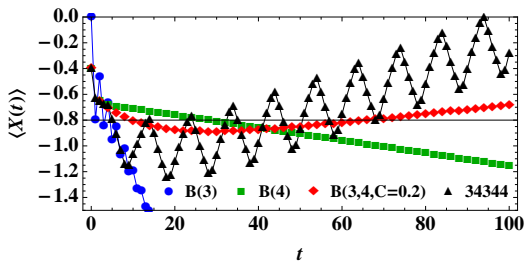}
\caption{\label{fig:capital_B34_34344}Expected capital $\langle X(t)\rangle$ for game $B(3),B(4),B(3,4,C=0.2)$ and switching sequence (34344). A line is added to $\langle X(t)\rangle$ for visualization purpose. For $B(4)$, $B(3,4,0.2)$ and the game sequence, every data point is a moving average over two consecutive time steps, in order to smooth out the oscillation. Parameters used: $p_b=0.1,p_g=0.67$. Both stochastic mixture and deterministic switching of game $B(3)$ and $B(4)$ could lead to a winning combined game, even without game A. }
\end{figure}

\makeatletter{}In the original Parrondo's game, while $M$ could be any integer larger than three, analysis of the game focused on the case of $M=3$. In early literature \cite{harmer_paradox_2000, pearce_parrondos_2000}, there were discussions on the effects of randomizing the parameter $M$. With preliminary results, Ref.~\cite{harmer_paradox_2000} demonstrated by randomizing $M$, additional complex and counter-intuitive phenomena could be observed. Inspired by this early effort, we systematically investigate the case where the value of $M$ of game B can take either $M_1$ or $M_2$ (without loss of generality we always assume $M_2>M_1$). In our work, we always assume that $p_{b1}=p_{b2},p_{g1}=p_{g2}$ for the two individual B games $B(M_1,p_{b1},p_{g1})$ and $B(M_2,p_{b2},p_{g2})$. Since $M$ is no longer a fixed value of the game, a game B with a particular value of $M$ is designated by $B(M)$. The full specification of a B game should be written as $B(M,p_b,p_g)$, but for simplicity we do not write its dependence on $p_b$ and $p_g$ explicitly. Similar to the mixture of game A and B in the original game, our extended model allowed the stochastic mixture and deterministic switching of $B(M_1)$ and $B(M_2)$. Using the notations of discrete time Markov chain, the stochastic mixture of $B(M_1)$ and $B(M_2)$ is equivalent to the linear combination of two transition matrices, $\Pi_B(M_1)$, $\Pi_B(M_2)$, corresponding to $B(M_1)$ and $B(M_2)$, respectively:
\begin{equation}
\Pi_B(M_1,M_2,C)=C\,\Pi_B(M_1)+(1-C)\Pi_B(M_2),
\end{equation}
where $C$ is the probability of using $B(M_1)$ in the stochastic mixture of $B(M_1)$ and $B(M_2)$ denoted by $B(M_1,M_2,C)$.
Notice that since the dimension of $\Pi_B(M_1,M_2,C)$ is $\textrm{LCM}(M_1,M_2)$ (LCM stands for least common multiple), Both transition matrices $\Pi_B(M_1)$ and $\Pi_B(M_2)$ have to be expanded to $\textrm{LCM}(M_1,M_2)\times\textrm{LCM}(M_1,M_2)$ matrices.  Fig.~\ref{fig:capital_B34_34344} shows that both stochastic mixture and deterministic switching could lead to a winning game.

\begin{figure}
\includegraphics[width=0.8\columnwidth]{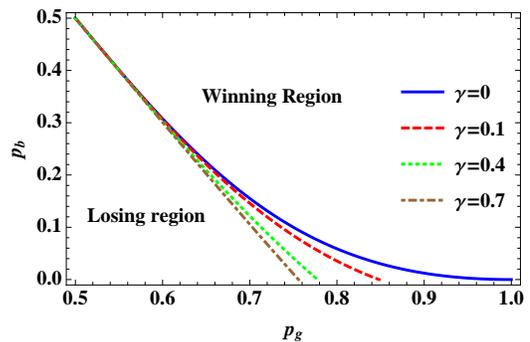}
\caption{\label{fig:parameter_space_mixing_paradox}
The winning and losing regions in the parameter space for the original Parrondo's game. By definition, the two regions are separated by Eq. \ref{eq:fair_original}. The parameter for game A, $p_A$, is fixed at 0.5 to preserve its analogy with pure diffusion process. As $\gamma$ increases, the fair game boundary becomes less and less convex.}
\end{figure}

We can also include game A into the stochastic mixture by 
\begin{equation}
\Pi=\gamma \Pi_A+(1-\gamma)\Pi_B(M_1,M_2,C).
\label{eq:mat_sto_mix_A_B_M1_M2}
\end{equation}

Markov chain analysis shows that given $C$ and $\gamma$, a Parrondo's game, be it individual game or stochastic mixture game, is wining, losing or fair depending only on the values of $p_b$ and $p_g$. A plot of ``winning-losing region" is particularly useful in explaining and investigating the seemingly paradoxical property of Parrondo's game. For the original Parrondo's game, a fair game corresponds to a point $(p_g,p_b)$ in the parameter space satisfying the following condition \cite{harmer_review_2002, pyke_random_2003} (for the derivation, please refer to the appendix):
\begin{equation}
\begin{gathered}
(\gamma p_A+(1-\gamma)p_b)\,(\gamma p_A+(1-\gamma)p_g)^2=\\
(1-(\gamma p_A+(1-\gamma)p_b))(1-(\gamma p_A+(1-\gamma)p_g))^2.
\end{gathered}\label{eq:fair_original}
\end{equation}

Eq.~\ref{eq:fair_original} defines the fair game boundary, in the parameter space $(p_g,p_b)$ and partitions the parameter space into a ``winning" region and a ``losing" region. See Fig.~\ref{fig:parameter_space_mixing_paradox} for the fair game boundary with several values of $\gamma$. The results should be interpreted in the following fashion: given a fixed value of $p_A$ (which is set to 0.5 to preserve the analogy with diffusion process) and $\gamma$, a particular Parrondo's game, corresponding to a point $(p_g,p_b)$, is winning if the point is above the fair game boundary (which is determined by the value of $\gamma$), losing if below the boundary.  When $\gamma$ increases, the fair game boundary shifts such that the winning region becomes larger. Also, when $\gamma$ increases, the fair game boundary becomes less convex, which will be an important factor when we consider the generalization to the extended game. 

\section{Strong Parrondo Effect in the Extended Model}

\makeatletter{}In the original Parrondo's game, the fair game condition (Eq.~\ref{eq:fair_original}) can be rewritten as
\begin{equation}
p_0p_1p_2=(1-p_0)(1-p_1)(1-p_2),
\end{equation}
where $p_i$ is the transition probability from state $i$ to state $i+1$, and implicitly we model the game as a discrete time Markov chain, in which the transition probability $P(i\rightarrow j)=0$ unless $j=i\pm1$ and $P(i\rightarrow j)=P(i+3\rightarrow j+3)$. In other words, it is a random walk with spatially-periodic transition probabilities. According to Ref.~\cite{pyke_random_2003}, a winning Parrondo's game corresponds to a Markov chain that is transient towards $\infty$, a fair game corresponds to a chain that is recurrent and a losing game corresponds to a chain that is transient towards $-\infty$. The fair game condition is therefore the condition under which the corresponding Markov chain is recurrent. For a random walk with spatially-periodic transition probabilities (period $L$), the condition under which it is recurrent, and therefore the fair game condition for a general Parrondo's game with period $L$, is
\begin{equation}
\prod_{i=0}^{L-1}p_i=\prod_{i=0}^{L-1}(1-p_i),
\end{equation}
where $p_i$ is the transition probability from state $i$ to state $i+1$. The fair game condition for $B(M_1,M_2)$ is therefore (please refer to the appendix for the derivation)
\begin{equation}
\begin{gathered}
p_{b}p_{g}^{Q}\alpha^{L/M_{1}-1}\beta^{L/M_{2}-1}=(1-p_{b})\\
(1-p_{g})^{Q}\left(1-\alpha\right)^{L/M_{1}-1}\left(1-\beta\right)^{L/M_{2}-1},
\end{gathered}
\label{eq:fair_game_condition_M1_M2}
\end{equation}
where $\alpha=\left(C\,p_b+(1-C)p_g\right)$, $\beta=\left((1-C)p_b+C\,p_g\right)$, $Q=L-L/M_{1}-L/M_{2}+1$, and $L=\textrm{LCM}(M_1,M_2)$. 

\begin{figure}[h]
\includegraphics[width=0.8\columnwidth]{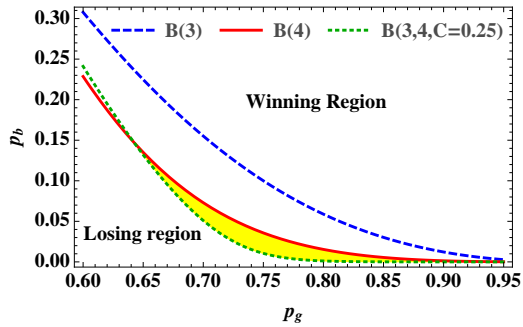}
\caption{\label{fig:phaseM34C025}
Fair game boundaries for $B(3),B(4)$ and $B(3,4,C=0.25)$. The shaded area is inside the winning region for $B(3,4,C=0.25)$ but is inside the losing regions for both $B(3)$ and $B(4)$. This means, given any point $(p_g,p_b)$ in the shaded area, stochastic mixture of $B(3)$ and $B(4)$ results in a winning combined game while the individual games are losing. We call the shaded region the strong Parrondo region, or more precisely the subset of parameter space, $\mathbb{S}(3,4,0.25)$.}
\end{figure}

Fig.~\ref{fig:phaseM34C025} shows the fair game boundaries for $B(3),B(4)$ and $B(3,4,C=0.25)$. The shaded area is of great interest: the area is inside the winning region for $B(3,4,C=0.25)$ but is also inside the losing regions for both $B(3)$ and $B(4)$. This means, given any point $(p_g,p_b)$ in the shaded area, stochastic mixture of $B(3)$ and $B(4)$ results in a winning combined game while the individual games are losing. In the extended model, two losing B games can be stochastically mixed to be a winning game, demonstrating strong Parrondo effect. We call the aforementioned area in the parameter space strong Parrondo region. More precisely, let us first define
\begin{equation}
\begin{gathered}
\mathbb{S}(M_1,M_2,C)\\
=\{(p_g,p_b)|g(M_1,M_2,C)>0\mbox{ and }g(M_1)<0\\
\mbox{ and }g(M_2)<0\}.
\end{gathered}
\label{eq:def_strong_Parrondo_region}
\end{equation}
This is the set of points in the parameter space where the combined game $B(M_1,M_2,C)$ has a positive long-term expected gain $g$, while the two individual games $B(M_1)$ and $B(M_2)$ have negative $g$. The set $\mathbb{S}(3,4,0.25)$ is thus the strong Parrondo region (shaded area in Fig.~\ref{fig:phaseM34C025}) when $C=0.25$. The statement that there exists a non-empty set $\mathbb{S}(M_1,M_2,C)$ for some value of $C$ is thus equivalent to the statement that strong Parrondo effect exist in the parameter space of $p_g$ and $p_b$ for this value of $C$.

However, not very pair of $B(M_1)$ and $B(M_2)$ is able to form a winning stochastic mixture game $B(M_1,M_2)$. One can show, using elementary geometry, that in the parameter space only when $\forall\,k\in\mathbb{N}, M_2\neq k\,M_1$, does $\mathbb{S}(M_1,M_2,C)$ exist for some value of $C$ in the range from 0 to 1. 

Here we only give a sketch of the proof. For games $B(M_1)$ and $B(M_2)$, the fair game boundaries are
\begin{equation}
p_b\,p_g^{M_1-1}=(1-p_b)(1-p_g)^{M_1-1}
\label{eq:fair_game_condition_M1}
\end{equation}
and
\begin{equation}
p_b\,p_g^{M_2-1}=(1-p_b)(1-p_g)^{M_2-1}.
\label{eq:fair_game_condition_M2}
\end{equation}

First we consider the end point at $p_g=0.5,p_b=0.5$, since Eq.~\ref{eq:fair_game_condition_M1_M2}, Eq.~\ref{eq:fair_game_condition_M1} and Eq.~\ref{eq:fair_game_condition_M2} will all pass this end point. We differentiate Eq.~\ref{eq:fair_game_condition_M1_M2},~\ref{eq:fair_game_condition_M1}~and~\ref{eq:fair_game_condition_M2} to obtain $\left.\mbox{d}p_b/\mbox{d}p_g\right|_{M_1,M_2}$, $\left.\mbox{d}p_b/\mbox{d}p_g\right|_{M_1}$ and $\left.\mbox{d}p_b/\mbox{d}p_g\right|_{M_2}$ at this point. The derivatives are 
\begin{equation}
\left.\dfrac{\mbox{d}p_b}{\mbox{d}p_g}\right|_{M}=1-M,
\end{equation}
so $\left.\mbox{d}p_b/\mbox{d}p_g\right|_{M_1}=1-M_1$ and $\left.\mbox{d}p_b/\mbox{d}p_g\right|_{M_2}=1-M_2$. For the stochastic mixture,  
\begin{equation}
\left.\dfrac{\mbox{d}p_b}{\mbox{d}p_g}\right|_{M_1,M_2}=1-\dfrac{M_1M_2}{C\,M_2+(1-C)M_1}.
\end{equation}
Clearly, $\left.\mbox{d}p_b/\mbox{d}p_g\right|_{M_2}<\left.\mbox{d}p_b/\mbox{d}p_g\right|_{M_1,M_2}<\left.\mbox{d}p_b/\mbox{d}p_g\right|_{M_1}$.
Next, we consider the other end point at $p_g=1,p_b=0$, since Eq.~\ref{eq:fair_game_condition_M1_M2}, Eq.~\ref{eq:fair_game_condition_M1} and Eq.~\ref{eq:fair_game_condition_M2} will also all pass through this end point.
Here, the derivatives $\mbox{d}p_b/\mbox{d}p_g$ are all zero for the three games, so instead we consider the three derivatives at $p_g=1-\epsilon$, where $\epsilon$ is a small number and has no relation with the parameters of the original Parrondo's game. At $p_g=1-\epsilon$,
\begin{equation}
\left.\dfrac{\mbox{d}p_b}{\mbox{d}p_g}\right|_{M_1,M_2}\sim\epsilon^{L-L/M_1-L/M_2}\quad\textrm{as }\epsilon\rightarrow0,
\end{equation}
and
\begin{equation}
\left.\dfrac{\mbox{d}p_b}{\mbox{d}p_g}\right|_{M}\sim\epsilon^{M-2}\quad\textrm{as }\epsilon\rightarrow0.
\end{equation}
 
When $\forall k\in\mathbb{N},M_2\neq k\,M_1$, $L-L/M_1-L/M_2>M_2-2>M_1-2$, which means $\left.\mbox{d}p_b/\mbox{d}p_g\right|_{M_1,M_2}$ goes to zero asymptotically faster than $\left.\mbox{d}p_b/\mbox{d}p_g\right|_{M_2}$ as $\epsilon\rightarrow0$. Considering that the slopes of the three fair game conditions are monotonically decreasing in $p_g$, there must exists a point of intersection between Eq.~\ref{eq:fair_game_condition_M1_M2} and Eq.~\ref{eq:fair_game_condition_M2}, and hence $\mathbb{S}(M_1,M_2,C)$ is non-empty for all $C\in(0,1)$.

If $\exists k\in\mathbb{N}$ such that $M_2=k\,M_1$, or equivalently, $M_2$ is a multiple of $M_1$, 
\begin{equation}
\left.\dfrac{\mbox{d}p_b}{\mbox{d}p_g}\right|_{M_1,M_2}\sim\epsilon^{M_2-k-1}\quad\textrm{as }\epsilon\rightarrow0.
\end{equation}

Clearly, $M_1-2<M_2-k-1<M_2-2$, which leads to the absence of intersection point between Eq.~\ref{eq:fair_game_condition_M1_M2} and Eq.~\ref{eq:fair_game_condition_M2}. For this reason, $\mathbb{S}(M_1,M_2,C)$ is empty for all $C\in(0,1)$. In other words, there is no strong Parrondo region in this case. For the case when $\forall k\in\mathbb{N},M_2\neq k\,M_1$, the position of the point of intersection $(p_g^*,p_b^*)$ can be calculated numerically in great accuracy. Since $(p_g^*,p_b^*)$ satisfies Eq.~\ref{eq:fair_game_condition_M2}, $p_g^*$ is sufficient to characterise the point of intersection. We use the notation $p_g^*(M_1,M_2,C)$ (or $p_b^*(M_1,M_2,C)$, since one is a function of the other) to designate the point of intersection as a function of $M_1,M_2$ and $C$. Numerical results show that in general $p_g^*(M_1,M_2,C)$ is an increasing function of $C$, while $p_b^*(M_1,M_2,C)$ is a decreasing function of $C$.

\begin{figure}
\centering
\subfloat[for $A(0.15)\oplus B(3)$, $A(0.15)\oplus B(4)$ and $A(0.15)\oplus B(3,4,0.2)$]
{\includegraphics[width=0.45\columnwidth]{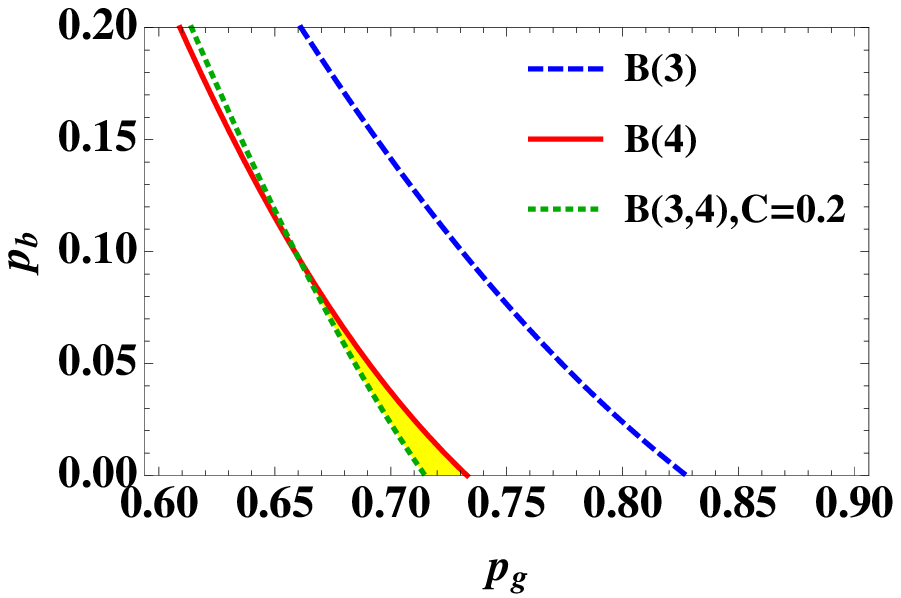}\label{fig:phaseM34C025g015}}
\subfloat[for $A(0.35)\oplus B(3)$, $A(0.35)\oplus B(4)$ and $A(0.35)\oplus B(3,4,0.2)$]
{\includegraphics[width=0.45\columnwidth]{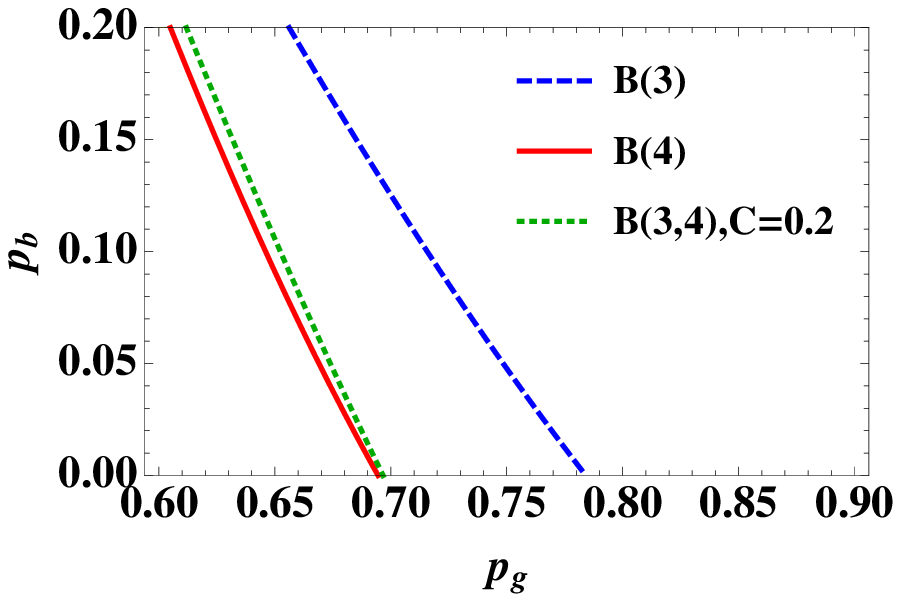}\label{fig:phaseM34C025g035}}
\caption{The effect of $B(3,4,0.2)$ stochastically mixed with game A with different values of $\gamma$. In \ref{fig:phaseM34C025g015}, despite that the fair game boundaries for the three games become less convex, the strong Parrondo region still exists, or $\mathbb{S}(3,4,0.2)$ is non-empty. In \ref{fig:phaseM34C025g035}, however, the strong Parrondo region ceases to exist, or equivalently $\mathbb{S}(3,4,0.2)=\emptyset$.}
\end{figure}

\begin{figure}
\includegraphics[width=0.8\columnwidth]{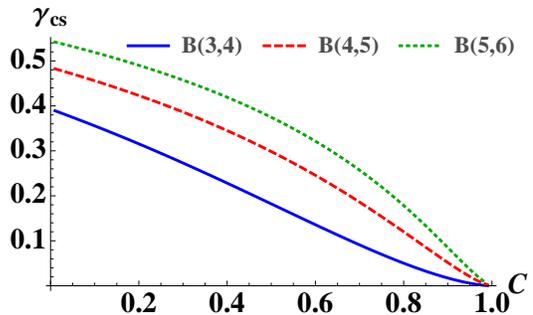}
\caption{\label{fig:gamma_c_vs_C}
Critical value of $\gamma$ for strong Parrondo effect, $\gamma_{cs}$, as a function of $C$ for several pairs of $M_1,M_2$. The value of $\gamma_{cs}$ is a decreasing function of $C$.}
\end{figure}

Physically, stochastic mixture of $B(M_1,M_2,C)$ with game A can be regarded as imposing a pure diffusion process with a particular strength on a random walk process in a spatially periodically fluctuating environment.
To see the effect of game A on the extended model, we can make a simple substitution on Eq.~\ref{eq:fair_game_condition_M1_M2} using $p_g\rightarrow \gamma\,p_A+(1-\gamma)p_g$ and $p_b\rightarrow\gamma\,p_A+(1-\gamma)p_b$. The matrix notation of this stochastic mixture is Eq.~\ref{eq:mat_sto_mix_A_B_M1_M2}. We use $A(\gamma)\oplus B(M_1,M_2,C)$ to designate such stochastic mixture. To preserve the analogy between game A and pure diffusion process, we set $p_A=0.5$. We show in Fig.~\ref{fig:phaseM34C025g015} and \ref{fig:phaseM34C025g035} for the effects of stochastic mixture with game A from the perspective of the fair game boundaries. As $\gamma$ increases, the fair game boundaries for B games become less convex in a way similar to the original Parrondo's game. Also, the point of intersection $p_g^*(M_1,M_2,C)$ moves downward in the parameter space. The upshot is the strong Parrondo region $\mathbb{S}(M_1,M_2,C)$ shrinks as $\gamma$ increases, and beyond a critical value $\gamma_{cs}$, $\mathbb{S}(M_1,M_2,C)$ becomes empty. More precisely,
\begin{equation}
\gamma_{cs}=\inf\left\{\gamma|\mathbb{S}(M_1,M_2,C)=\emptyset\right\}.
\end{equation}
$\gamma_{cs}$ can be calculated numerically in great accuracy. Since when $p_b^*(M_1,M_2,C)=0$, $\gamma=\gamma_{cs}$, we can calculate $\gamma_{cs}$ by solving $p_b^*(M_1,M_2,C)=0$. See Fig.~\ref{fig:gamma_c_vs_C} for $\gamma_{cs}$ as a function of $C$ for several pairs of $M_1,M_2$. In general, $\gamma_{cs}$ is a decreasing function of $C$. We can understand this result in the following way: since $p_b^*(M_1,M_2,C)$ is a decreasing function of $C$, as $C$ increases, less amount of game A is needed to ``drag" the point of intersection down to $p_b=0$.
Numerical calculation shows that the addition of game A will not introduce strong Parrondo region to the parameter space when $\exists\,k$ such that $M_2=k\,M_1$.

\section{Weak Parrondo effect in the extended model}

\makeatletter{}The investigation of the weak Parrondo effect in the extended model requires the calculation of the expected gain (Eq.~\ref{eq:expected_gain}). This can be achieved in more than one way. One can solve Eq.~\ref{eq:mat_sto_mix_A_B_M1_M2} for the stationary probability vector and obtain $g$ from it. One can also derive a general formula for $g$ like the one in Ref.~\cite{amengual_paradoxical_2006,derrida_velocity_1983}.  
We extend the definition of the strong Parrondo region to accommodate the weak Parrondo region: 
\begin{equation}
\begin{gathered}
\mathbb{S}(M_1,M_2,C,g_0)\\
=\{(p_g,p_b)|g(M_1,M_2,C)>g_0\mbox{ and }g(M_1)<g_0\\
\mbox{ and }g(M_2)<g_0\},
\end{gathered}
\end{equation}
which is the set of points in the parameter space where the combined game $B(M_1,M_2,C)$ with long-term expected gain more than $g_0$ while the two individual games $B(M_1)$ and $B(M_2)$ gain less than $g_0$ per time step. The strong Parrondo region is thus a special case with $g_0=0$, or equivalently $\mathbb{S}(M_1,M_2,C)\equiv\mathbb{S}(M_1,M_2,C,0)$. The weak Parrondo region, or
\begin{equation}
\begin{gathered}
\mathbb{W}(M_1,M_2,C)\\
=\{(p_g,p_b)|g(M_1,M_2,C)>g(M_1)\mbox{ and }\\
g(M_1,M_2,C)>g(M_2)\}
\end{gathered}
\end{equation}
satisfies
\begin{equation}
\mathbb{W}(M_1,M_2,C)=\bigcup_{g_0\in\mathbb{R}}\mathbb{S}(M_1,M_2,C,g_0).
\end{equation}

\begin{figure}
\includegraphics[width=0.8\columnwidth]{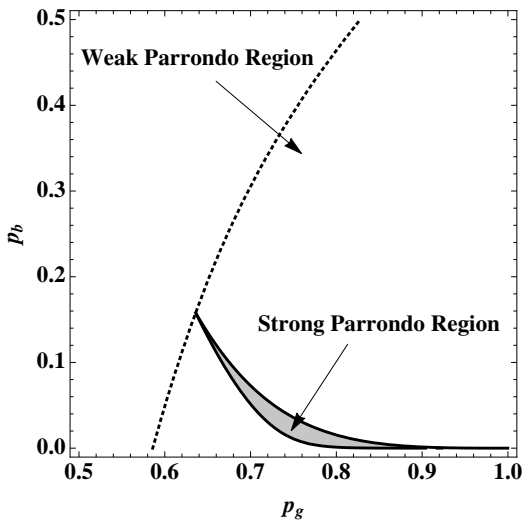}
\caption{\label{fig:weak_strong_Parrondo_BM34_C02}
Weak Parrondo region $\mathbb{W}(3,4,0.2)$ (to the right of the dotted line) and strong Parrondo region $\mathbb{S}(3,4,0.2)$ (shaded region). The set $\mathbb{S}(3,4,0.2)$ is a non-trivial subset of $\mathbb{W}(3,4,0.2)$. In fact, under this setting $(M_1=3,M_2=4,C=0.2)$, $\mathbb{W}(3,4,0.2)$ is much larger in area than $\mathbb{S}(3,4,0.2)$.
}
\end{figure}

\begin{figure}
\includegraphics[width=0.8\columnwidth]{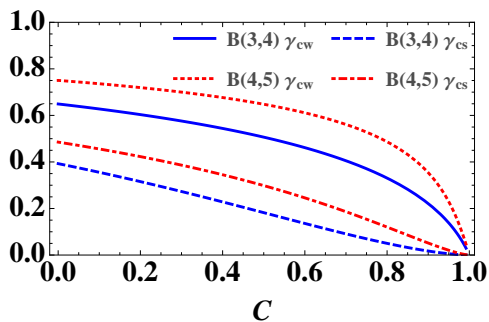}
\caption{\label{fig:gamma_cs_cw_vs_C_BM34_BM45}
The critical value of $\gamma$ for the weak Parrondo region, $\gamma_{cw}$, and $\gamma_{cs}$ as a function of $C$ for $B(3,4)$ and $B(3,5)$. For every $C$, $\gamma_{cw}>\gamma_{cs}$.
}
\end{figure}

Fig.~\ref{fig:weak_strong_Parrondo_BM34_C02} shows the weak Parrondo region $\mathbb{W}(3,4,0.2)$ and strong Parrondo region $\mathbb{S}(3,4,0.2)$. Clearly, $\mathbb{S}(3,4,0.2)$ is a non-trivial subset of $\mathbb{W}(3,4,0.2)$. 

We have just shown that the further addition of game A could shrink the strong Parrondo region to an empty set. Numerical results show that the weak Parrondo region $\mathbb{W}(M_1,M_2,C)$ also shrink as $\gamma$ increases and the property $\mathbb{S}(M_1,M_2,C)\subset\mathbb{W}(M_1,M_2,C)$ holds. There also exists a critical value $\gamma_{cw}(C)$ (in general larger than $\gamma_{cs}(C)$), beyond which the weak Parrondo region becomes an empty set. See Fig.~\ref{fig:gamma_cs_cw_vs_C_BM34_BM45} for $\gamma_{cw}$ and $\gamma_{cs}$ as a function of $C$ for $B(3,4)$ and $B(3,5)$. We observe that $\gamma_{cw}>\gamma_{cs}$ for every $C$. Therefore, backed up by numerical calculations, weak Parrondo effect is a generalization of strong Parrondo effect, in the parameter space, in the sense that $\mathbb{S}(M_1,M_2,C)\subset\mathbb{W}(M_1,M_2,C)$. The weak Parrondo effect is more robust than strong Parrondo effect in the face of imposed diffusion process. 

\section{Extended Parrondo's Game as a Discrete Brownian Ratchet}

\makeatletter{}While Parrondo’s games were originally inspired by the flashing ratchet, no direct relation was established between them until the work of Allison et al. \cite{allison_physical_2002-1,allison_aspects_2009-1} and Toral et al. \cite{toral_parrondos_2003,toral_fokker-planck_2003,amengual_paradoxical_2006} appeared. The establishment of the connection requires the discretization of the Fokker-Planck equation and the matching up of the discrete Fokker-Planck equation with the master equation of Parrondo's game. In this work, we adopt the discretization scheme employed by Toral et al since it produces intuitive discrete ratchet potential and probability current. Nevertheless, it can be shown that the potentials resulted from the two approaches coincide in the limit of an infinitesimally small space-discretized step\cite{amengual_discretetime_2004}. The discrete ratchet potential corresponding to a Parrondo's game is:
\begin{equation}
V_i=-\dfrac{1}{2}\ln\left[\prod_{k=1}^i\dfrac{q_{k-1}}{1-q_k}\right],
\label{ratchet_potential}
\end{equation}
where $q_k$ is the transition probability from state $k$ to $k+1$. The ratchet potential corresponding to a winning game is decreasing in trend. Similarly, the ratchet potential corresponding to a losing game is increasing in trend and for fair game the ratchet potential is constant in trend.  To capture the trend of a ratchet potential, we define a quantity called the macroscopic bias
\begin{equation}
E=-\dfrac{V_L}{L}=\dfrac{1}{2L}\ln\left[\prod_{k=1}^L\dfrac{q_k}{1-q_k}\right],
\label{macroscopic_bias}
\end{equation}
which is just the average potential drop over one spatial period. Because of the minus sign, a winning game has a positive bias, a losing game has a negative bias and a fair game has zero bias, which is consistent with the fair game condition for the Parrondo's game. Thus, the quantity $\prod_{k=1}^L\dfrac{q_k}{1-q_k}$ in Eq.~\ref{macroscopic_bias} provides a convenient way for relating the result of the Parrondo's game to the potential drop in the Brownian ratchet.

\begin{figure}[h]
\includegraphics[width=0.8\columnwidth]{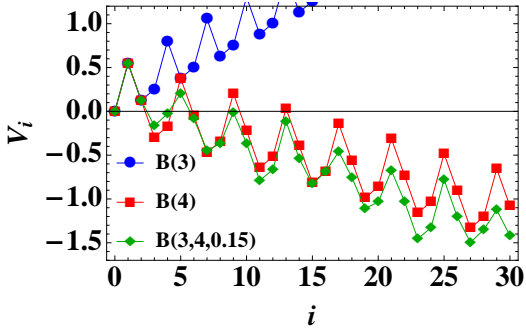}
\caption{\label{fig:V_discrete_BM34_C015}
Discrete ratchet potential for $B(3)$, $B(4)$ and $B(3,4,0.15)$. The parameters are $p_b=0.1$ and $p_g=0.7$.
}
\end{figure}

In the discrete ratchet picture, the original Parrondo's paradox is equivalent to the situation that two ratchet potentials with zero macroscopic bias, or even slightly negative bias, combining through Eq. \ref{ratchet_potential}, form a ratchet potential with positive macroscopic bias. In the original Parrondo's game, starting from a discrete ratchet potential $V_i(M)$ corresponding to fair game B (zero macroscopic bias), the addition of game A will introduce a positive bias and modify the intensity of local fluctuations (measured by $V_i-i\,E$). In the extended Parrondo's game, however, the combination of two ratchet potentials $V_i(M_1)$ and $V_i(M_2)$ leads to a complicated ratchet potential $V_i(M_1,M_2,C)$ which in general has different macroscopic bias from $V_i(M_1)$ and $V_i(M_2)$, different intensity of local fluctuations, and vastly different overall potential profile, as shown in Fig.~\ref{fig:V_discrete_BM34_C015}.

The probability current \cite{amengual_exact_2004} is 
\begin{equation}
J=\dfrac{1}{2}P_0^{\textrm{st}}\dfrac{1-e^{2V_L}}{\sum_{j=1}^L\dfrac{e^{2V_j}}{2-2p_j}},
\label{eq:prob_curr}
\end{equation}
where $P_0^{\textrm{st}}$ is the stationary probability at state 0 modulo $L$ given by the implicit relation
\begin{equation}
P_i^{\textrm{st}}=e^{-2V_i}\left[P_0^{\textrm{st}}-2J\sum_{j=1}^i\dfrac{e^{2V_j}}{2-2p_j}\right]
\end{equation}
and the normalization condition, $\sum_{i=0}^{L-1}P_i^{\textrm{st}}=1$, gives a complicated solution of $P_0^{\textrm{st}}$. This formulation of probability current is consistent with the result of Markov chain analysis since $g(p_g,p_b,M_1,M_2,C)=\textrm{LCM}(M_1,M_2)J(p_g,p_b,M_1,M_2,C)$ analytically.

While the sign of $E$ tells whether a Parrondo's game is winning or losing (positive $E$ corresponds to a winning game), the relative magnitudes of two ratchet potentials $E_1$, $E_2$ do not tell which one is winning more, i.e. having a larger $g$, since there is no one-to-one correspondence between $E$ and $g$. In fact, it can be shown that
\begin{equation}
\begin{gathered}
\max_{0\le p_g\le 1,0\le p_b\le1}g(p_g,p_b,M_1,M_2,C)=\dfrac{2}{e^{-2E_0}+1}-1
\\ \mbox{ subject to } E(p_g,p_b)=E_0.
\end{gathered}
\end{equation} 
The maximum is achieved when $p_g=p_b$, which corresponds to the case where $V_i=-\dfrac{i}{2}\ln\left(\dfrac{p_g}{1-p_g}\right)$. The minimum of $g(p_g,p_b,M_1,M_2,C)$ subject to $E(p_g,p_b)=E_0$ is zero, when $p_g\rightarrow1,p_b\rightarrow0$ and $V_1\rightarrow\infty$, corresponding to the case when $V_1$ is so large that it blocks the movement of particles entirely, regardless of whether macroscopic bias is finite and non-zero.

Let us use the following notation
\begin{equation}
\begin{gathered}
\mathbb{D}(M_1,M_2,C)\\
=\{(p_g,p_b)|E(M_1,M_2,C)>E(M_2)\},
\end{gathered}
\end{equation}
\begin{equation}
\begin{gathered}
\mathbb{F}(M_1,M_2,C)\\
=\{(p_g,p_b)|E(M_1,M_2,C)>E(M_1)\}
\end{gathered}
\end{equation}
and
\begin{equation}
\mathbb{P}=\{(p_g,p_b)|0\le p_b\le0.5,0.5\le p_g\le1\}.
\end{equation}
Since the sign of $E$ determine whether a game is winning or losing, the relative magnitudes of $E$ for different games could give a naive expectation of whether one game performs better (measured in the long-term average gain $g$) than the other. In the application to the extended game, it provides a simple guideline in predicting whether the combined game $B(M_1,M_2,C)$ performs better than the two individual games. It is not difficult to show that $E(M_2)>E(M_1)$ if $M_2>M_1$ for $0\le p_b\le0.5$ and $0.5\le p_g\le1$. Also, $E(M_1)<E(M_1,M_2,C)<E(M_2)$ if $M_2$ is a multiple of $M_1$ for $0\le p_b\le0.5$ and $0.5\le p_g\le1$. These two properties coincide with the real game performances. For other $M$ pairs, the boundary is described by 
\begin{equation}
\begin{gathered}
p_g^R\alpha^{L/M_1-1}\beta^{L/M_2-1}=(1-p_g)^R\\
(1-\alpha)^{L/M_1-1}(1-\beta)^{L/M_2-1},
\end{gathered}
\label{eq:equal_bias_condition}
\end{equation}
where $R=L-L/M_1-L/M_2-M_2+2$, $\alpha=C\,p_b+(1-C)p_g$, $\beta=((1-C)p_b+C\,p_g)$, $L=\textrm{LCM}(M_1,M_2)$. One can verify that the point of intersection between Eq.~\ref{eq:fair_game_condition_M1_M2} and Eq.~\ref{eq:fair_game_condition_M2} is a solution to Eq.~\ref{eq:equal_bias_condition}, which means the existence of the point of intersection guarantee the existence of $\mathbb{D}(M_1,M_2,C)$. Since for $M_2\neq k\,M_1\quad\forall k\in\mathbb{N}$, the point of intersection exists for all $C$, $\mathbb{D}(M_1,M_2,C)$ exists for all $C$. The existence of $\mathbb{F}(M_1,M_2,C)$, however, can only be found out numerically.

\begin{figure}[h]
\centering
\subfloat[for $B(3)$, $B(4)$ and $B(3,4,0.2)$]
{\includegraphics[width=0.45\columnwidth]{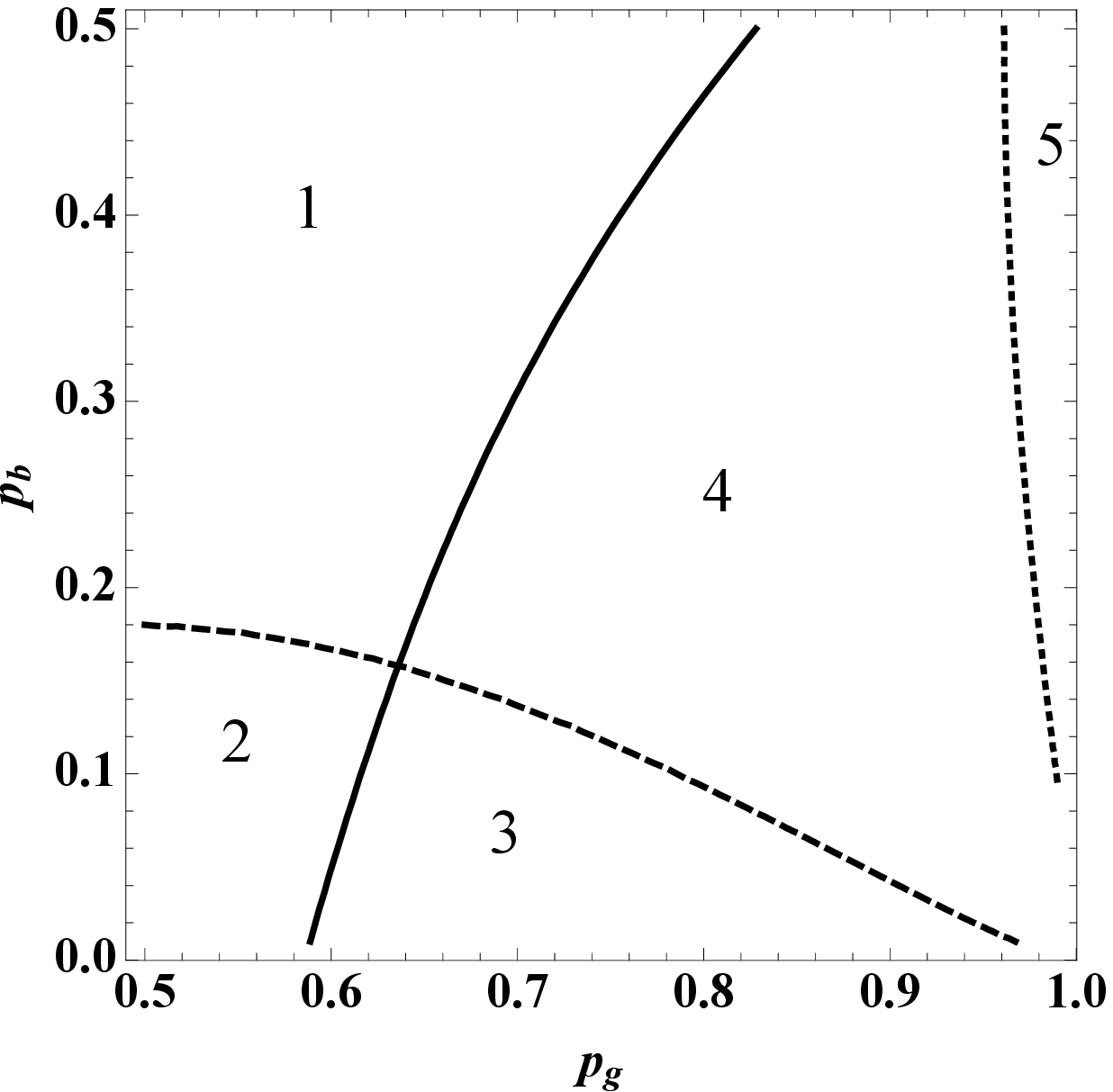}\label{fig:BM34_C02_parameter_space}}
\subfloat[for $B(3)$, $B(4)$ and $B(3,4,0.7)$]
{\includegraphics[width=0.45\columnwidth]{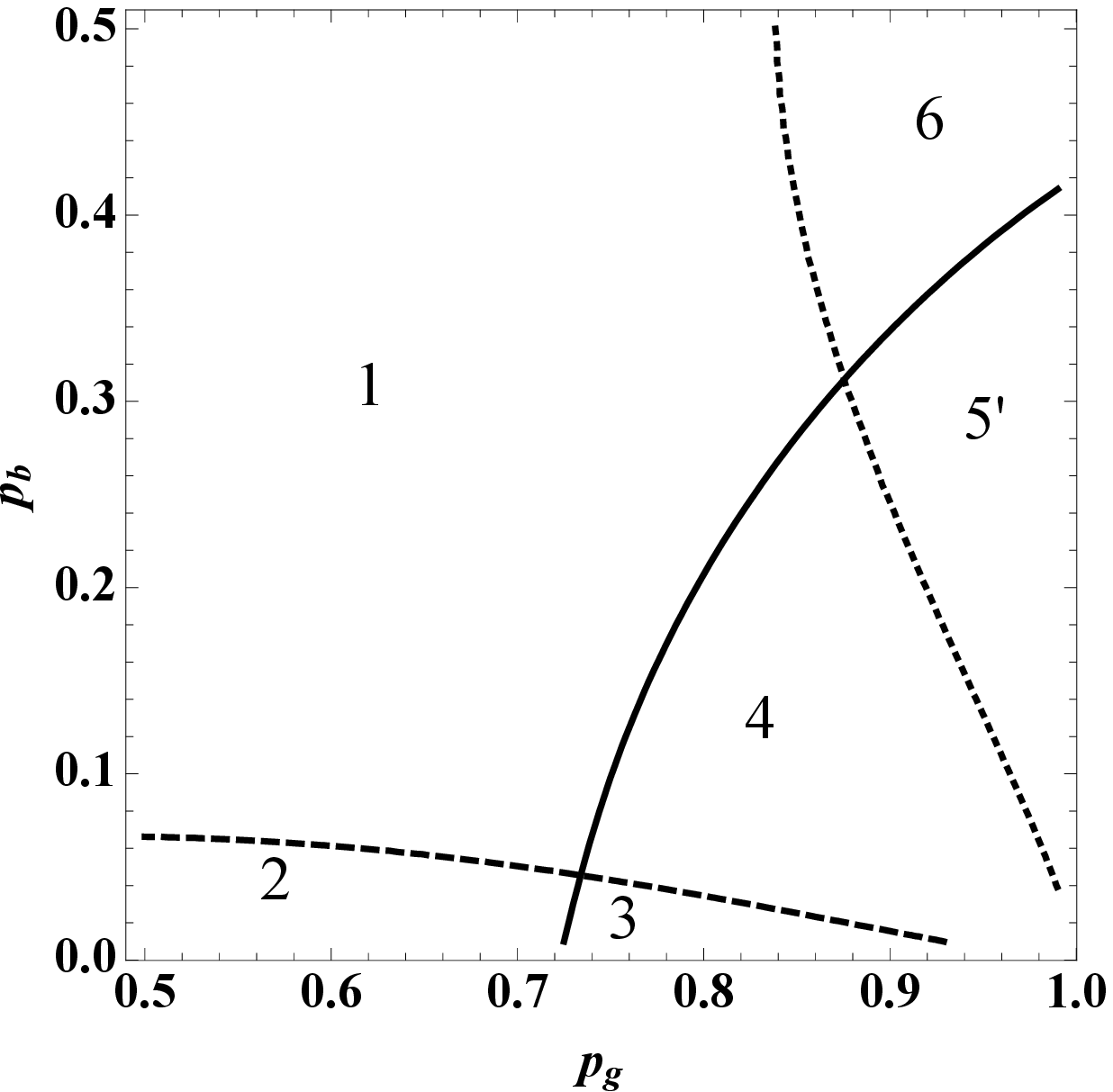}\label{fig:BM34_C07_parameter_space}}
\caption{Various partitions in the parameter space. \protect\subref{fig:BM34_C02_parameter_space} for small $C$ and \protect\subref{fig:BM34_C07_parameter_space} for large $C$. Set $2\cup3$ is $\mathbb{D}(3,4,C)$. Set $1\cup2\cup3\cup4$ is $\mathbb{F}(3,4,C)$. In \protect\subref{fig:BM34_C02_parameter_space}, set $3\cup4\cup5$ is $\mathbb{W}(3,4,C)$. In \protect\subref{fig:BM34_C07_parameter_space}, set $3\cup4\cup5'$ is $\mathbb{W}(3,4,C)$. Set 1 is where $E(3)<E(3,4,C)<E(4)$ and $g(3)<g(3,4,C)<g(4)$. Set 3 is where $E(3)<E(4)<E(3,4,C)$ and $g(3)<g(4)<g(3,4,C)$. Set $1\cup3$ is where relation among the bias of the three games is consistent with the relation among the game performance of the three games.}
\label{fig:BM34_parameter_space}
\end{figure}

We show in Fig.~\ref{fig:BM34_parameter_space} the partitions of parameter space with regard to relative magnitude of macroscopic bias and long-term expected gain of the three games. In the region of parameter space defined by the set 1, $E(3)<E(3,4,C)<E(4)$ and $g(3)<g(3,4,C)<g(4)$ hold simultaneously. In set 3 of the parameter space, $E(3)<E(4)<E(3,4,C)$ and $g(3)<g(4)<g(3,4,C)$ hold simultaneously. Therefore, in these two sets, the relation among the bias of the three games is consistent with the relation among the game performances of the three games. 

To summarise, when $M_2$ is a multiple of $M_1$, the relation among $E(M_1),E(M_2)$ and $E(M_1,M_2,C)$ is consistent with the relation among $g(M_1)$, $g(M_2)$ and $g(M_1,M_2,C)$. In other words, $E(M_1)<E(M_1,M_2,C)<E(M_2)$ and $g(M_1)<g(M_1,M_2,C)<g(M_2)$ hold simultaneously. When $\forall k\in\mathbb{N},M_2\neq k\,M_1$, there exists one subset $\mathbb{F}(M_1,M_2,C)\backslash(\mathbb{W}(M_1,M_2,C)\cup\mathbb{D}(M_1,M_2,C))$ in which $E(3)<E(3,4,C)<E(4)$ and $g(3)<g(3,4,C)<g(4)$ hold simultaneously and another subset $\mathbb{W}(M_1,M_2,C)\cap\mathbb{D}(M_1,M_2,C)$ in which $E(3)<E(4)<E(3,4,C)$ and $g(3)<g(4)<g(3,4,C)$ hold simultaneously. The exact counterpart of game performance in the discrete ratchet picture is the probability current (Eq.~\ref{eq:prob_curr}), which has a different form from $E$, so it is not surprising to see the two quantities behave differently on some occasions. The interesting thing is in some region of the parameter space, namely $1\cup3$ in Fig.~\ref{fig:BM34_parameter_space}, the two quantities behave in a similar way. 

\section{Conclusions}

\makeatletter{}We have extended the original Parrondo's game to allow $M$ to be either $M_1$ or $M_2$. We have discussed the distinction between the strong Parrondo effect and the weak Parrondo effect, which plays an important role in the extended model. In the extended game, two individual games $B(M_1)$ and $B(M_2)$ can stochastically mix to form a better-performing game, including both the strong and the weak Parrondo effects, if and only if $M_2$ is not a multiple of $M_1$. If $M_2$ is a multiple of $M_1$, it is impossible to obtain a better-performing game by combining $B(M_1)$ and $B(M_2)$, meaning the absence of both strong and weak Parrondo effect. Further addition of game A, in analogy to a pure diffusion process imposed on the game, can destroy the strong and the weak Parrondo effect, but the weak Parrondo effect is more robust against the imposed pure diffusion process.

We have shown the physical meaning of our extended model in terms of discrete ratchet potentials obtained through the discretization of Fokker-Planck equation. We have identified an important quantity, the macroscopic bias, or the average drop in the discrete ratchet potential in one spatial period. While macroscopic bias is not the same as the performance of the game, measured by long-term expected gain, the relation among the bias of the three games ($B(M_1)$, $B(M_2)$ and $B(M_1,M_2,C)$) in some case is in agreement with the relative performance of the three games. If $M_2$ is a multiple of $M_1$, the relation among $E(M_1)$, $E(M_2)$ and $E(M_1,M_2,C)$ is in agreement with the relation among $g(M_1)$, $g(M_2)$ and $g(M_1,M_2,C)$ in the whole parameter space. If $M_2$ is not a multiple of $M_1$, there exists a proper subset in which the relations of the two quantities agree with each other. 

Our model assumes $p_{b1}=p_{b2}$, $p_{g1}=p_{g2}$ for $B(M_1,p_{b1},p_{g1})$ and $B(M_2,p_{b2},p_{g2})$ for the benefit of a systematic investigation, since if we remove these assumptions, the parameter space will be four-dimensional. More features are expected to emerge if we relax these constrictions. For future work, we can allow $M$ to be one of three possible values, such as 3,4 and 5. A stochastic mixture of $B(3)$, $B(4)$ and $B(5)$ is expected to produce a game that performs better than the stochastic mixture of any pair chosen from $B(3)$, $B(4)$ and $B(5)$ for some point in the parameter space. Moreover, Toral's game \cite{toral_capital_2002} is one version of multi-player Parrondo's game, and it contains a modified game A that redistributes the wealth between players, resembling strong interaction between Brownian particles, from the perspective of discrete ratchet. Incorporating Toral's modified game A into our extended model, we could create a new version of Parrondo's game which is the counterpart of solitonic flashing ratchet \cite{floria_transport_2002}.  

\section{Acknowledgement}

K.Y. Szeto acknowledges the support of grant FS-GRF13SC25 and FS-GRF14SC28.  We also thank the referee for pointing out several important references that we do not know, especially Ref.~\cite{harmer_brownian_2001} which has suggested several open questions, one of which is the varying M problem.

\section{Appendix: the derivation of fair game condition}

\makeatletter{}\begin{figure}[h]
\centering
\subfloat[a general random walk]
{\scalebox{0.8}{
\includegraphics[]{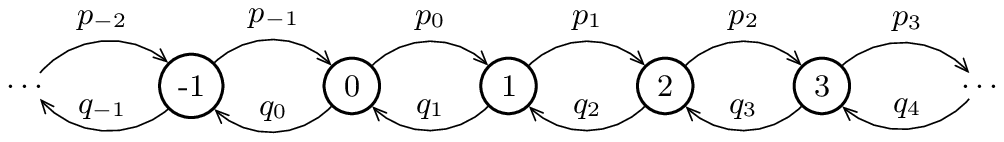}
\label{fig:inf_walk_general}}}\\
\subfloat[game $B(M=3)$]
{\scalebox{0.8}{
\includegraphics[]{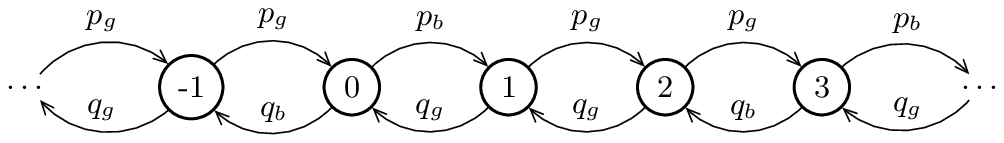}
\label{fig:inf_walk_BM3}}}
\caption{The discrete time Markov chain for \ref{fig:inf_walk_general} general random walk \ref{fig:inf_walk_BM3} for game $B(M=3)$. In \ref{fig:inf_walk_general} $p_i=P(i\rightarrow i+1)$ is the transition probability from state $i$ to $i+1$, and $q_i\equiv1-p_i$. In \ref{fig:inf_walk_BM3}, $p_{3k}=p_b, p_{1+3k}=p_{2+3k}=p_g,\forall k\in\mathbb{Z}$. $q_g\equiv1-p_g$ and $q_b\equiv1-p_b$.}
\label{fig:inf_walk}
\end{figure}

\begin{figure}
\scalebox{0.8}{
\includegraphics[]{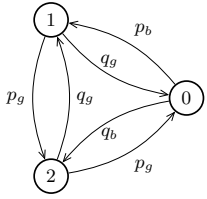}
}
\caption{Reduced Markov chain for game $B(M=3)$.}
\label{fig:ring_walk_BM3}
\end{figure}

\begin{figure}
\subfloat[extended $B(3)$]
{
\scalebox{0.8}{
\includegraphics[]{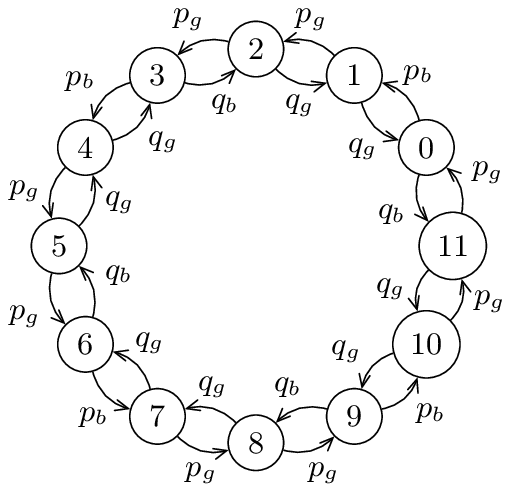}
}\label{fig:ring_extended_BM3}
}
\subfloat[extended $B(4)$]
{
\scalebox{0.8}{
\includegraphics[]{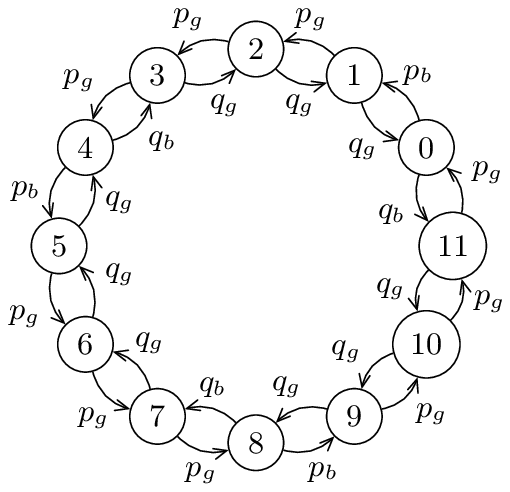}
}\label{fig:ring_extended_BM4}
}
\caption{Extended finite state Markov chain for \ref{fig:ring_extended_BM3} $B(3)$ and \ref{fig:ring_extended_BM4} $B(4)$.}
\end{figure}

Parrondo's game, be it original or extended, can be modeled as a discrete time random walk on the integer set $\mathbb{Z}$ (Fig. \ref{fig:inf_walk_general}). For example, game $B(M=3)$ has a corresponding Markov chain depicted in Fig. \ref{fig:inf_walk_BM3}. The Markov chain for game A is a special case of game B where $p_g=p_b$. The stochastic mixture of game A and B will not modify the structure of the Markov chain, but will merely change the transition probabilities at each site by $p_g\rightarrow\gamma p_A+(1-\gamma)p_g$ and $p_b\rightarrow\gamma p_A+(1-\gamma)p_b$. According to Ref.~\cite{nowak_two_2005}, in the context of Parrondo's game, a fair game will have a corresponding recurrent Markov chain. For a winning game, the chain is transient towards $\infty$ while for a losing game the chain is transient towards $-\infty$. For a random walk which transition probability is periodic, i.e. $p_i=p_{i+kL},\forall k\in\mathbb{Z}$, the condition of recurrence is
\begin{equation}
\prod_{i=0}^{L-1}p_i=\prod_{i=0}^{L-1}(1-p_i),
\label{eq:recurr_condition}
\end{equation}
where $L=3$ for $B(M=3)$. The proof of the condition for recurrence can be found in many standard textbook on Markov chain and will be omitted here. Plug in $p_0=\gamma p_A+(1-\gamma)p_b$ and $p_1=p_2=\gamma p_A+(1-\gamma)p_g$ we will recover Eq. \ref{eq:fair_original}.

Because of the periodic nature of the transition probabilities, Markov chain depicted in Fig. \ref{fig:inf_walk_BM3} can be reduced to one with only three states, depicted in Fig. \ref{fig:ring_walk_BM3}. However, the state probability becomes
\begin{equation}
\hat{\pi_i}(t)=\sum_{k=-\infty}^{\infty}\pi_{i+kL},
\end{equation}
where $L=3$ for $B(M=3)$. 

For a extended game $B(M_1,M_2,C)$, say $M_1=3$ and $M_2=4$, first we have to extend the finite state Markov chains with three states (for $B(3)$) and four states (for $B(4)$) to two equivalent Markov chains with twelve states (Fig. \ref{fig:ring_extended_BM3} and \ref{fig:ring_extended_BM4}). To get the Markov chain corresponding to $B(3,4,C)$, we need to add the transition probabilities together by $p_i=C\,p_i(M=3)+(1-C)p_i(M=4)$. In $B(3,4,C)$, the transition probability can only be one of the following four values: $p_g$, $p_b$, $C\,p_g+(1-C)p_b$ and $C\,p_b+(1-C)p_g$. Out of the twelve transition probabilities $\{p_i\}$,
\begin{equation}
\begin{gathered}
p_0=p_b,\\
p_3=p_6=p_9=C\,p_b+(1-C)p_g,\\
p_4=p_8=C\,p_g+(1-C)p_b,\\
p_1=p_2=p_5=p_7=p_{10}=p_{11}=p_g.
\end{gathered}
\end{equation}
Plug them into the recurrence condition Eq. \ref{eq:recurr_condition} and one will get
\begin{equation}
\begin{gathered}
p_{b}p_{g}^6\alpha^{3}\beta^{2}=(1-p_{b})
(1-p_{g})^{6}\left(1-\alpha\right)^{3}\left(1-\beta\right)^{2},
\end{gathered}
\end{equation}
where $\alpha=\left(C\,p_b+(1-C)p_g\right)$, $\beta=\left((1-C)p_b+C\,p_g\right)$. This is a special case for fair game condition for general pair of $M_1$ and $M_2$, i.e. Eq. \ref{eq:fair_game_condition_M1_M2}.

\bibliography{./my_lib}

\end{document}